\documentclass[pra,twocolumn,amsmath,amssymb,superscriptaddress,longbibliography]{revtex4-2}

\usepackage{graphicx}% Include figure files
\usepackage{dcolumn}% Align table columns on decimal point
\usepackage{bm}% bold math
\usepackage{times}
\usepackage{xcolor}
\definecolor{darkblue}{rgb}{0, 0, 0.8}
\usepackage[colorlinks=true, breaklinks=true, linkcolor=darkblue, citecolor=darkblue, urlcolor=darkblue]{hyperref}

\usepackage{braket}
\newcommand{\quota}[1]{``#1''} % to put quotations use \quota{the text}, it is like writing ``the text''

\begin{document}
\title{Benchmarking direct and indirect dipolar spin-exchange interactions\\ between two Rydberg atoms}

\author{Gabriel~Emperauger}
\affiliation{Université Paris-Saclay, Institut d'Optique Graduate School, CNRS, Laboratoire Charles Fabry, 91127 Palaiseau Cedex, France}

\author{Mu~Qiao}
\affiliation{Université Paris-Saclay, Institut d'Optique Graduate School, CNRS, Laboratoire Charles Fabry, 91127 Palaiseau Cedex, France}

\author{Guillaume~Bornet}
\affiliation{Université Paris-Saclay, Institut d'Optique Graduate School, CNRS, Laboratoire Charles Fabry, 91127 Palaiseau Cedex, France}

\author{Cheng~Chen}
\affiliation{
Institute of Physics, Chinese Academy of Sciences, Beijing 100190, China
}
\affiliation{Université Paris-Saclay, Institut d'Optique Graduate School, CNRS, Laboratoire Charles Fabry, 91127 Palaiseau Cedex, France}

\author{Romain~Martin}
\affiliation{Université Paris-Saclay, Institut d'Optique Graduate School, CNRS, Laboratoire Charles Fabry, 91127 Palaiseau Cedex, France}

\author{Yuki~Torii~Chew}
\affiliation{Université Paris-Saclay, Institut d'Optique Graduate School, CNRS, Laboratoire Charles Fabry, 91127 Palaiseau Cedex, France}
\affiliation{Institute for Molecular Science, National Institutes of Natural Sciences, Okazaki 444-8585, Japan}

\author{Bastien~G\'ely}
\affiliation{Université Paris-Saclay, Institut d'Optique Graduate School, CNRS, Laboratoire Charles Fabry, 91127 Palaiseau Cedex, France}

\author{Lukas~Klein}
\affiliation{Université Paris-Saclay, Institut d'Optique Graduate School, CNRS, Laboratoire Charles Fabry, 91127 Palaiseau Cedex, France}

\author{Daniel~Barredo}
\affiliation{Université Paris-Saclay, Institut d'Optique Graduate School, CNRS, Laboratoire Charles Fabry, 91127 Palaiseau Cedex, France}
\affiliation{Nanomaterials and Nanotechnology Research Center (CINN-CSIC), Universidad de Oviedo (UO), Principado de Asturias, 33940 El Entrego, Spain}

\author{Antoine~Browaeys}
\affiliation{Université Paris-Saclay, Institut d'Optique Graduate School, CNRS, Laboratoire Charles Fabry, 91127 Palaiseau Cedex, France}

\author{Thierry~Lahaye}
\affiliation{Université Paris-Saclay, Institut d'Optique Graduate School, CNRS, Laboratoire Charles Fabry, 91127 Palaiseau Cedex, France}

\date{\today}

\begin{abstract}
We report on the experimental characterization of various types of spin-exchange interactions between two individual atoms, where pseudo-spin degrees of freedom are encoded in different Rydberg states. For the case of the direct dipole-dipole interaction between states of opposite parity, such as between $\ket{nS}$ and $\ket{nP}$, we investigate the effects of positional disorder arising from the residual atomic motion, on the coherence of spin-exchange oscillations. We then characterize an indirect dipolar spin exchange, i.e., the off-diagonal part of the van der Waals effective Hamiltonian that couples the states $\ket{nS}$ and $\ket{(n+1)S}$. 
Finally, we report on the observation of a new type of dipolar coupling, made resonant using addressable light-shifts and involving four different Rydberg levels: this exchange process is akin to electrically induced F\"orster resonance, but featuring local control. It exhibits an angular dependence distinct from the usual $1-3\cos^2(\theta)$ form of the resonant dipolar spin-exchange.
\end{abstract}

\maketitle

Arrays of Rydberg atoms are now one of the leading platforms for quantum science and technology. They have allowed for several breakthroughs in quantum computing~\cite{Bluvstein2022,Bluvstein2024,Reichardt2024,Tsai2024}, and have proven to be a very versatile tool for analog quantum simulation of spin models~\cite{Browaeys2020}. One of the key reasons for this success is the strong, controllable, dipolar interaction between pairs of Rydberg atoms. Measuring accurately the interactions and their coherence time is thus essential for improving the platform and ensuring a faithful implementation of many-body problems of interest~\cite{browaeys_experimental_2016}. Such benchmarks have been realized in the context of ultrafast Rydberg interactions~\cite{Chew_2022}, Rydberg-dressing~\cite{Kim_Heisenberg_2024} and polar molecules~\cite{Holland_2023,bao_dipolar_2023,Ruttley_2025,Picard_2025} and pointed to the crucial role of positional disorder as a source of dephasing in quench experiments. Here, we aim at performing this benchmark for typical experimental parameters used in current intermediate-scale quantum simulations with Rydberg arrays~\cite{Bornet2023, Chen2023, Emperauger2025, Qiao2025}.

\begin{figure}%[b!]
\centering
\includegraphics{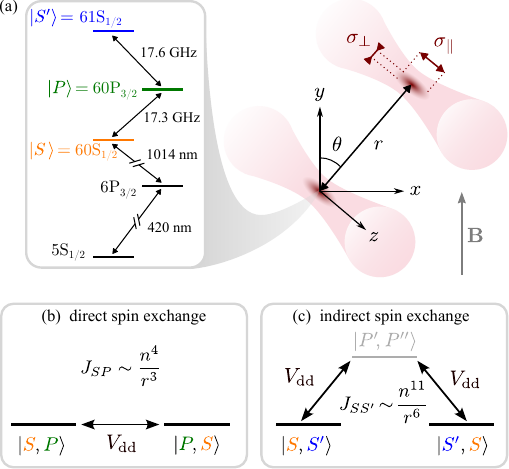}
\caption{{\bf Spin-exchange experiments.}
    (a)~Sketch of a pair of atoms interacting at an average distance~$r$, forming an angle~$\theta$ with the magnetic field~$\mathbf{B}$. Fluctuations of the atomic positions are represented by the dark red regions, with standard deviations $\sigma_\perp$ along the radial direction of the tweezers ($x$ and $y$) and $\sigma_\parallel$ along the axial direction ($z$). Inset: relevant energy levels of a rubidium atom, where possible spin-encoding states used in this work are colored.
    Spin exchange can arise from direct~(b) or indirect~(c) dipolar interaction, giving rise to different scalings of the coupling constant $J$ with distance $r$ and principal quantum number $n$.
    }
\label{fig:fig1}
\end{figure}

Two Rydberg atoms separated by a distance~$r$ much larger than the size of the electronic wavefunctions [Fig.~\ref{fig:fig1}(a)] interact predominantly via the electric dipole-dipole Hamiltonian
\begin{equation}
 \hat{V}_{\rm dd}=\frac{1}{4\pi\varepsilon_0}\frac{\hat{\mathbf{d}}_1\cdot\hat{\mathbf{d}}_2-3(\hat{\mathbf{d}}_1\cdot\mathbf{e}_r)(\hat{\mathbf{d}}_2\cdot \mathbf{e}_r)}{r^3},
    \label{eq:vdd}
\end{equation}
where $\hat{\mathbf{d}}_i$ stands for the electric dipole of atom $i$ and $\mathbf{e}_r$ is the unit vector along the interatomic axis. For two atoms prepared in the same Rydberg level, the effect of $\hat{V}_{\rm dd}$ appears only in second-order perturbation theory~\cite{cohen-tannoudji_atom-photon_1998}, giving rise to the van der Waals shift which is used for quantum simulation of the transverse-field Ising model~\cite{Bernien2017,Lienhard2018,Scholl2021,Ebadi2021,Semeghini2021}. However, if one prepares the two atoms in two distinct Rydberg levels that are dipole-coupled, such as for instance $\ket{nS}$ and $\ket{nP}$, the pair state $\ket{nS,nP}$ is directly coupled by $\hat{V}_{\rm dd}$ to $\ket{nP,nS}$, which has the same energy: this resonant dipole-dipole interaction gives rise to a coherent exchange of the internal states of the two atoms, with a strength scaling as $1/r^3$ [Fig.~\ref{fig:fig1}(b)]. Encoding a pseudo-spin $1/2$ in those two states implements the dipolar XY model, which has already been studied in several experiments~\cite{Barredo2015,deLeseleuc2019,Scholl2022,Chen2023a,Bornet2023,Bornet2024,Chen2023,Emperauger2025}. The spin can also be encoded in Rydberg levels with the same parity, such as $\ket{nS}$ and $\ket{n'S}$, leading to an indirect spin-exchange interaction between $\ket{nS, n'S}$ and $\ket{n'S, nS}$, that scales as $1/r^6$ [Fig.~\ref{fig:fig1}(c)]. Such mapping has been used to realize quantum simulations of XXZ models in disordered ensembles~\cite{Signoles_2021,franz_observation_2024}. Combining the direct and indirect spin exchanges with three encoding Rydberg levels enables the study of doped magnetism~\cite{Qiao2025} or spin-1 physics~\cite{Mogerle2024,Liu2024}.

In this article, we experimentally investigate several aspects of the dipolar spin-exchange interaction, which we define as the effective coupling between $\ket{\uparrow, \downarrow}$ and $\ket{\downarrow, \uparrow}$ where $\ket{\uparrow}$ and $\ket{\downarrow}$ are two Rydberg levels. 
First, we study the influence on the direct spin-exchange of the residual motion of the atoms, that leads to fluctuations in the interatomic distance and thus in the interaction strength. We find a good agreement between our measurements, that use two different methods, and numerical simulations taking into account independently measured experimental imperfections. Second, we extend this study to the indirect spin-exchange
that arises when encoding our spin-$1/2$ degrees of freedom in 
$\ket{nS}$ and $\ket{(n+1)S}$ [Fig.~\ref{fig:fig1}(c)].
Thanks to the agreement between the simulations and the data, we can analyze the contributions of each imperfection and provide a roadmap for future improvements of the platform. 
Finally, we demonstrate that, by using an extra light-shifting beam that brings two pair states involving four Zeeman states into resonance, one implements an exotic dipolar coupling term with an unusual angular dependence.

\section{Experimental setup}
Our experimental setup has been described elsewhere~\cite{Bornet2024}. We trap two individual rubidium atoms ($^{87}$Rb) in optical tweezers at a controlled distance~$r\sim 10$-40~µm, and optically pump them into the hyperfine ground state $\ket{5S_{1/2},F=2,m_F=2}$. We then switch off the tweezers and excite both atoms to the same Rydberg state $\ket{S} \equiv \ket{60S_{1/2},m_J=1/2}$ with a stimulated Raman adiabatic passage (STIRAP) using two lasers at 420~nm and 1014~nm. We then use a local addressing laser and a global microwave field~\cite{Bornet2024} to prepare the targeted initial state ---which depends on the type of experiment. After that, we let the atoms evolve in free flight during an interacting time~$t$, under the dipole-dipole interaction~$\hat{V}_{\rm dd}$. Finally, we read out the internal state of each atom by selectively deexciting the state~$\ket{S}$ to the ground state manifold $5S_{1/2}$~\footnote{The deexcitation pulse is rather long (a few hundreds of ns) compared with the typical evolution time under the dipole-dipole interactions. To limit the effect of interactions during the measurement, we perform a fast ($\sim 50$~ns) \quota{freezing pulse} before the deexcitation pulse, that transfers the population of the non-imaged Rydberg state ($\ket{P}$ or $\ket{S'}$) to other Rydberg states that do not interact with~$\ket{S}$. Additional details on this pulse can be found in the Supplementary Material of~\cite{Qiao2025} or in~\cite{Emperauger_PhD_2025}.}, switching the tweezers back on and performing fluorescence imaging. Imaged atoms are interpreted as being in the state~$\ket{S}$, whereas atoms that are in another Rydberg state are expelled from the tweezers by the ponderomotive force. During the Rydberg sequence, we set the quantization axis by a magnetic field $|\mathbf{B}|\approx46$~G along~$y$ or along~$z$, in order to isolate two $m_J$-states. This also allows us to control the angle $\theta$ between the interatomic axis $\boldsymbol{n}$ and the magnetic field~$\mathbf{B}$, by tuning the positions of the tweezers in the $xy$ plane [Fig.~\ref{fig:fig1}(a)].

Before the Rydberg sequence, we use Raman sideband cooling~\cite{kaufman_cooling_2012,thompson_coherence_2013} to reduce the thermal fluctuations of atomic motion inside the tweezers, achieving typical average occupation numbers for the motional states of $\bar{n}_\perp \sim 0.5$ radially, and $\bar{n}_\parallel \sim 10$ axially. The corresponding temperatures are $T_\perp \sim 4$~µK and $T_\parallel \sim 10$~µK. The standard deviation of the position along a direction~$i$ depends on the tweezer trapping frequency~$\omega_i$ as $\sigma_i = \sqrt{\hbar ( \bar{n}_i + 1/2) / m \omega_i}$, and the standard deviation of the velocity distribution as $\sigma_{v_i} = \omega_i \sigma_i$. The total position uncertainty after a duration~$t$ of free flight is thus $\sigma_i(t) = \sqrt{\sigma_i^2 + (\sigma_{v_i} t)^2}$, as pictured in Fig.~\ref{fig:fig1}(a).
Before the Rydberg sequence, we adiabatically reduce the trapping depth by a factor $\zeta \in [0.001,1]$, and thus the trapping frequencies by a factor $\sqrt{\zeta}$ (starting from $\omega_\perp \sim 2\pi \times 100$~kHz and $\omega_\parallel \sim 2\pi \times 20$~kHz). 
This corresponds to a positional disorder of $\sigma_\perp \sim [30,200]$~nm, $\sigma_\parallel \sim [250,1400]$~nm, $\sigma_{v_\perp} \sim [4,20]$~nm$/$µs and $\sigma_{v_\parallel} \sim [6,30]$~nm$/$µs, leading in turn to fluctuations of the interatomic distance. To first order in $\sigma_\perp$ and $\sigma_\parallel$, the standard deviation of distances is given by $\Delta r (t) = \sqrt{2} \sigma_\perp(t)$, which is a few percents of the average distance~$r$.

\begin{figure*}%[b!]
\centering
\includegraphics{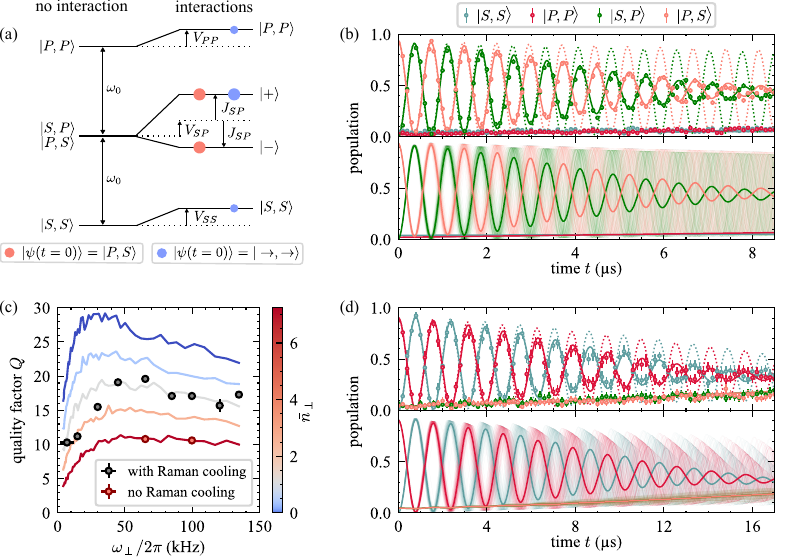}
\caption{{\bf Benchmarking a direct dipolar spin-exchange.} (a)~Two-atom energy levels without (left) and with (right) the dipole-dipole interactions~$\hat{V}_{\rm dd}$, showing the initial populations in a spin-exchange experiment (pink disks) and in a two-atom Ramsey experiment (blue disks). (b)~Top: time evolution of the populations of the four pair states, at distance $r=14.9$~µm, $\theta=90$° and with initial trapping frequency~$\omega_\perp \sim 2\pi \times 10$~kHz, for $\ket{P} = \ket{60P_{3/2},m_J=3/2}$: experimental data (points with error bars denoting the standard error on the mean), simulations including (solid line) and excluding (dotted lines) atomic motion. Bottom: illustration of how averaging over many individual oscillations for different interatomic distances leads to damping.
(c)~Evolution of the quality factor of a spin-exchange experiment with the initial radial trapping frequency~$\omega_\perp$, which is controlled by the trap depth of the optical tweezers before switching them off (the atoms being in free flight during the Rydberg sequence). Experimental parameters: $r=17$~µm, $\theta=90$°, $\ket{P} = \ket{60P_{3/2},m_J=1/2}$. The black points (resp. red points) indicate the data with (resp. without) Raman sideband cooling. Lines are classical simulations taking into account various experimental imperfections (see text). The color encodes the radial motional state $\bar{n}_\perp$ (and $\bar{n}_\parallel \sim 10$). The longitudinal trapping frequency~$\omega_\parallel$ varies proportionally to~$\omega_\perp$, but it is not represented since the effect of longitudinal disorder on the quality factor is negligible.
(d)~Same as (b) but for a ``two-atom'' Ramsey experiment (see text).
}
\label{fig:fig2}
\end{figure*}

\section{Direct dipolar spin exchange}
Here, we focus on the resonant dipole-dipole interaction that couples the pair states $\ket{S,P}$ and $\ket{P,S}$ with $\ket{P} = \ket{60P_{3/2},m_J=3/2}$. The dipole-dipole interaction~$\hat{V}_{\rm dd}$ restricted to the basis $\{\ket{S,S}, \ket{S,P}, \allowbreak \ket{P,S}, \ket{P,P}\}$ leads to the following Hamiltonian:
\begin{align}
	H(\mathbf{r}) = \hbar
	\begin{bmatrix}
		V_{S S} & 0 & 0 & 0 \\
		0 & V_{S P} & J_{S P} & 0 \\
		0 & J_{S P} & V_{S P} & 0 \\
		0 & 0 & 0 & V_{P P}
	\end{bmatrix}
	\label{eq:H_eff_two_levels}
\end{align}
where all diagonal terms are van der Waals shifts that scale as $1/r^6$, and the  off-diagonal term scales as $1/r^3$. Such a Hamiltonian can be mapped onto an XXZ model using~$\ket{\uparrow} = \ket{S}$ and $\ket{\downarrow} = \ket{P}$~\cite{franz_observation_2024}. The associated eigenstates are: $\ket{S,S}$, with eigenenergy~$\hbar V_{SS}$; $\ket{\pm} = (\ket{S,P} \pm \ket{P,S})/\sqrt{2}$, with eigenenergies~$\hbar (V_{SP} \pm J_{SP})$; and $\ket{P,P}$, with eigenenergy~$\hbar V_{PP}$ as illustrated in Fig.~\ref{fig:fig2}(a). For an interatomic distance~$r=14.9$~µm, $J_{S P}$ dominates all other terms by more than one order of magnitude. This is the regime we operated in to implement the dipolar XY model in previous works \cite{deLeseleuc2019,Scholl2022,Chen2023a,Bornet2023,Bornet2024,Chen2023,Emperauger2025}.

We want to measure the interaction energy $J_{S P}$ and probe the effects of various experimental imperfections on the dynamics of the system, compared with the ideal case of perfectly localized atoms evolving unitarily under the effective Hamiltonian of Eq.~(\ref{eq:H_eff_two_levels}). For that, we use two complementary methods, which we call \quota{spin exchange} and \quota{2-atom Ramsey experiment}, and benchmark them against numerical simulations that account for calibrated experimental imperfections.

We first consider the \quota{spin exchange} method~\cite{Barredo2015}. We initialize the atoms in $\ket{P,S}$ to prepare the symmetric superposition of $\ket{+}$ and $\ket{-}$ [pink disks in Fig.~\ref{fig:fig2}(a)], in which case we expect out-of-phase oscillations at frequency $2J_{SP}$ for the populations of the states $\ket{P,S}$ and $\ket{S,P}$. In this method, the oscillation frequency is directly the interaction energy (up to a factor 2) and does not depend on the van der Waals terms; however, it does not give access to the sign of $J_{SP}$. Figure~\ref{fig:fig2}(b) shows the experimental results. We observe long-lived oscillations that allow us to extract the interaction energy $|J_{SP}| = 2\pi \times (668 \pm 1)$~kHz, in agreement with ab-initio calculations using the software PairInteraction~\cite{Weber2017}. The fit of the oscillations by a sine wave with a Gaussian envelope $\exp(-t^2 / 2 \tau^2)$ yields a damping time $\tau = 3.8 \pm 0.1$~µs corresponding to a quality factor $Q = \tau |J_{SP}| = 16 \pm 0.4$.

To model our experimental results quantitatively, we perform numerical simulations including various imperfections: (i) \emph{Positional disorder} gives rise to fluctuations of the interaction energies.
(ii) \emph{Finite Rydberg lifetimes} create leakage from the ideal two-level picture. Each of the Rydberg levels $\ket{S}$ and $\ket{P}$ has two decay channels: it can decay to the ground state $5S_{1/2}$ due to spontaneous emission~\footnote{Actually, the level $\ket{S}$ does not decay directly to the ground state $\ket{g} = 5S_{1/2}$, but it decays to low-lying $P$ levels that all end up quickly in $\ket{g}$.}, with respective rates $\Gamma_S^{0K}=($260~µs$)^{-1}$ and $\Gamma_P^{0K}=($472~µs$)^{-1}$; or to other Rydberg states due to black-body radiation in our room temperature setup, with respective rates $\Gamma_S^\mathrm{BB}=($157~µs$)^{-1}$ and $\Gamma_P^\mathrm{BB}=($161~µs$)^{-1}$. 
(iii) \emph{Preparation errors} limit the initial contrast. The Rydberg excitation pulse leaves a small atomic population of~$2$~\% in the ground state, and local microwave rotations also have a finite infidelity of about 2~\%.
(iv) \emph{Detection errors} reduce the overall contrast of the oscillations. An atom in $\ket{S}$ (resp. in $\ket{P}$) has a finite probability of about $2$~\% to be detected in $\ket{P}$ (resp. in $\ket{S}$)~\cite{deLeseleuc_Barredo_2018}.

The numerical simulations work as follows. We first draw initial positions and velocities for the two atoms according to a thermal distribution. We then numerically solve the Lindblad equation, taking into account effects (ii) to (iv). We finally average the results over typically 1000 realizations of the shot-to-shot positional disorder [effect (i)], considering that the atoms are in free flight. The assumption that the atoms are in free flight is only justified at early times, before the van der Waals forces affect the atomic positions (see App.~\ref{App:vdW_forces}). The resulting populations are shown as thick solid lines in Fig.~\ref{fig:fig2}(b), whereas each realization corresponds to a thin line. Averaging these oscillations, which have slightly different frequencies due to positional disorder, gives rise to the observed damping. Our error model allows us to fully capture the experimental data, including the contrast of the oscillations, using as free parameters $\bar{n}_\perp \simeq 0.3$ and $\bar{n}_\parallel \simeq 11$ (these values are in reasonable agreement with those obtained by Raman sideband spectroscopy).

To further test the influence of atomic motion onto the damping of the spin-exchange, we perform another set of measurements using $\ket{P} = \ket{60P_{3/2},m_J=3/2}$ and a distance~$r=17$~µm, where we scan the initial position and velocity dispersions by changing the depth of the optical tweezers by a factor~$\zeta$ before the free flight. The extracted quality factors are plotted in Fig.~\ref{fig:fig2}(c) as a function of the initial radial trapping frequency~$\omega_\perp \propto \sqrt{\zeta}$ (the effect of longitudinal positional disorder on $\Delta r$ being negligible in this regime of parameters). We observe an optimum around~$\omega_\perp\sim 2\pi \times 50$~kHz, where $Q$ reaches 20. Numerical simulations with the same model [solid lines in Fig.~\ref{fig:fig2}(c)] predict a similar behavior; we obtain the best agreement for $\bar{n}_\perp = 1.1 \pm 0.3$ [gray solid line], slightly higher than previously quoted for this new set of measurements. The optimum originates from a trade-off between the initial position dispersion~$\sigma_\perp \propto 1/\sqrt{\omega_\perp}$ and the initial velocity dispersion~$\sigma_{v_\perp} \propto \sqrt{\omega_\perp}$: at low trapping frequencies, $\sigma_\perp$ is the major contribution to the damping, whereas at high trapping frequencies~$\sigma_{v_\perp}$ dominates. In App.~\ref{App:positional_disorder}, we design a model to better understand this behavior; we also study the case of trapped Rydberg atoms and show that it could significantly improve the damping for trapping frequencies~$\omega_\perp > 2\pi \times 30$~kHz.

Another method for measuring the dipole-dipole interactions is an adaptation of a Ramsey sequence to the case of two atoms. Starting from $\ket{S,S}$ after the global Rydberg excitation, we apply a global microwave $\pi/2$ pulse to prepare the state $\ket{\rightarrow,\rightarrow}$ where $\ket{\rightarrow}\equiv \left(\ket{S} + \ket{P}\right) / \sqrt{2}$ corresponds to a spin along~$x$. We let the system evolve for a time $t$ and apply another global microwave $\pi/2$ pulse to measure in the $x$ basis. Unlike above, this sequence does not require local addressing, which can induce additional decoherence (reduced lifetime for the addressed state~$\ket{S}$, momentum kicks due to the ponderomotive force). However, contrary to a spin-exchange experiment, the initial state gets decomposed over three eigenstates of $H$, each one with a different eigenenergy: $\ket{\rightarrow,\rightarrow} = \frac{1}{\sqrt{2}} \left[\ket{+} + \frac{1}{\sqrt{2}}\left(\ket{S,S} + \ket{P,P}\right)\right]$ [blue disks in Fig.~\ref{fig:fig2}(a)]. In the absence of a careful phase compensation, this would lead to a beating instead of a simple oscillation. This beating can be compensated by a simple microwave detuning $\delta = (V_{PP} - V_{SS})/2$, such that the accumulated phase for the two states $\ket{S,S}$ and $\ket{P,P}$ is the same. An alternative method would be to use a dynamical decoupling technique, such as done in~\cite{bao_dipolar_2023,Holland_2023}. After the second $\pi/2$-pulse, one expects an oscillation between the states $\ket{S,S}$ and $\ket{P,P}$ with an angular frequency as $\tilde{J}_{SP} \equiv J_{SP} + V_{SP} - (V_{SS}+V_{PP})/2$. As a result, this method informs about the relative sign between $J_{SP}$ and van der Waals terms.

Experimentally, we optimize the microwave detuning $\delta$ by maximizing the contrast of the oscillation, resulting in the curve shown in Fig.~\ref{fig:fig2}(d). We obtain $|\tilde{J}_{SP}| = 2\pi \times (635 \pm 1)$~kHz, in agreement with the calculated interaction energies $J_{SP} = -2\pi \times 675$~kHz, $V_{SS} = 2\pi \times 12$~kHz, $V_{SP} = 2\pi \times 17$~kHz and $V_{PP} = -2\pi \times 52$~kHz. The initial contrast and the damping of the oscillation are again well explained by simulations with the same experimental imperfections as above. Here, the damping arises mainly from positional disorder; Rydberg lifetimes contribute to a lesser extent~\footnote{In addition, a slight overcompensation of the detuning, which was experimentally set to $\delta = -2\pi \times 40$~kHz whereas the ideal compensation should happen for $\delta = -2\pi \times 32$~kHz according to the simulations, leads to a small rise up of the probabilities of the states $\ket{S,P}$ and $\ket{P,S}$ at long times due to the above-mentioned beating phenomenon, mimicking additional damping.}. Compared with the case of the spin exchange with the same experimental parameters [Fig.~\ref{fig:fig2}(b)], the damping time is twice as long, but the oscillation frequency is also twice as small, so that the quality factor is approximately the same ($Q = 13.8 \pm 0.2$).

\begin{figure}%[t]
\centering
\includegraphics{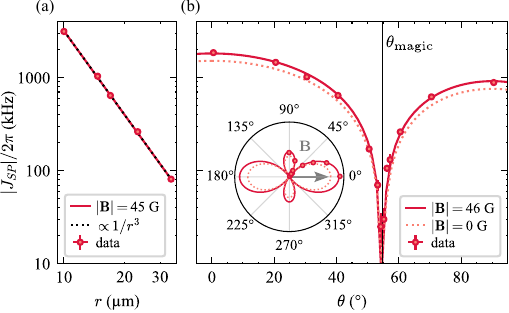}
\caption{{\bf Characterization of a direct dipolar spin-exchange.}
(a)~Scaling of the measured coupling with interatomic distance, showing the expected~$1/r^3$ dependence, for a fixed angle~$\theta=90$° and for $\ket{P} = \ket{60P_{3/2},m_J=1/2}$.
(b)~Measured angular dependence of the dipolar spin coupling (circles) in a semi-log plot, for a distance~$r=9.9$~µm and for $\ket{P} = \ket{60P_{3/2},m_J=-1/2}$. The solid (resp. dotted) lines give the theoretical value in the presence of a magnetic field of 46~G (resp. at zero field). The vertical black line indicates the theoretical angle~$\theta_{\rm magic}=\arccos(1/\sqrt{3})$ at which $J_{SP}$ theoretically vanishes. The inset shows the same data in polar coordinates.}
\label{fig:fig3}
\end{figure}

Using the calibrated method of the spin exchange, we now measure the dependence of $|J_{SP}|$ on the average interatomic distance~$r$, using $\ket{P} = \ket{60P_{3/2},m_J=1/2}$. The obtained interaction energies [Fig.~\ref{fig:fig3}(a)] agree perfectly with the theory prediction~\cite{Weber2017}, and show the expected~$1/r^3$ behavior. We also measure $|J_{SP}|$ as a function of the angle $\theta$ between the quantization and the interatomic axis. To do so, we set the magnetic field $\mathbf{B}$ to be in the atomic plane with $|\mathbf{B}|\approx 46$~G. For this set of measurements, we use $\ket{P} = \ket{60P_{3/2},m_J=-1/2}$ at a distance~$r=9.9$~µm; this does not affect the results apart from an overall multiplication of $|J_{SP}|$ by a factor~$3.6$. The results are summarized in Fig.~\ref{fig:fig3}(b) and display the characteristic dipolar pattern $1-3\cos^2(\theta)$; they match very well with the theory prediction~\cite{Weber2017}, shown as a solid line. An important parameter to take into account for the accuracy of the theory prediction is the value of the magnetic field, which mixes the fine structure of the $60P$ manifold~\footnote{In our case, the state $\ket{P}$ is redefined as $\ket{\tilde{P}} = c \ket{P} + c' \ket{P'}$ with $\ket{P} = \ket{60P_{3/2},m_J=-1/2}$ and $\ket{P'} = \ket{60P_{1/2},m_J=-1/2}$. A numerical simulation using~\cite{Weber2017} gives $c \approx 0.997$ and $c' \approx -0.073$. The dipole-dipole coupling becomes $J_{S\tilde{P}} = \bra{S,\tilde{P}} V_\mathrm{dd} \ket{\tilde{P},S}/\hbar = |c|^2 \bra{S,P} V_\mathrm{dd} \ket{P,S}/\hbar + 2 \Re(c^* c') \bra{S,P} V_\mathrm{dd} \ket{P',S}/\hbar + O\left(|c'|^2\right)$. As a result, the coupling in the presence of the magnetic field increases by about $20$~\% compared with the case with no mixing, $J_{SP} = \bra{S,P} V_\mathrm{dd} \ket{P,S}/\hbar$ [dotted line in Fig.~\ref{fig:fig3}(e)]. Note that this mixing does not affect the angular dependence of $J_{S\tilde{P}}$, since all terms involve a $d^+ d^-$ transition and thus have the same angular dependence $\propto 1-3\cos^2(\theta)$.}.

\begin{figure}%[t]
\centering
\includegraphics{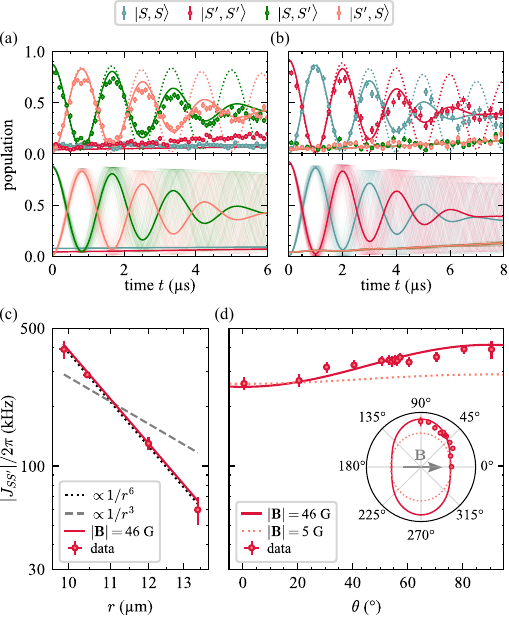}
\caption{{\bf Benchmarking and characterization of an indirect dipolar spin-exchange.} (a,b)~Coherent spin-exchange oscillations between the states $\ket{S,S'}$ and $\ket{S',S}$ using the same two methods and rendering as in Fig.~\ref{fig:fig2}. (c) Scaling of the measured coupling with interatomic distance, clearly showing its $1/r^6$ dependence, for a fixed angle $\theta=90$°. The dotted and dashed lines are fits by power laws with fixed power-law exponents (respectively 6 and 3). (d) Measured (circles) and calculated (lines) angular dependence of the coupling at~$r=9.9$~µm, showing a weak anisotropy.}
\label{fig:fig4}
\end{figure}

\section{Indirect dipolar spin exchange}

We now benchmark the indirect dipolar spin exchange, that couples the pair states $\ket{S,S'}$ and $\ket{S',S}$, using $\ket{S'} = \ket{61S_{1/2},m_J=1/2}$. To measure this off-diagonal interaction~$|J_{S,S'}|$, we use the same methods as for the direct dipole coupling, since both cases are described by the same effective Hamiltonian [Eq.~(\ref{eq:H_eff_two_levels})]; only the dependence with distance and with the angle differs. The transition between $\ket{S}$ and $\ket{S'}$ is addressed by a microwave two-photon transition through the intermediate states of the $60P$ manifold, with a detuning of about $200$~MHz from $\ket{60P_{3/2},m_J=1/2}$ and an effective Rabi frequency of 10~MHz.
Figure~\ref{fig:fig4}(a) shows a spin-exchange oscillation at~$r=10.4$~µm that is correctly reproduced by the numerical simulations, using an interaction energy $J_{SS'} = 2\pi \times 299$~kHz, $\bar{n}_\perp = 1.7$ and $\bar{n}_\parallel = 11$. The obtained value of~$\bar{n}_\perp$ is higher than in the previous experiments, but it is compatible with that of a direct spin-exchange oscillation performed in the same experimental conditions; we attribute this to a reduced cooling efficiency. Here, the effect of positional disorder is more detrimental than in the direct spin exchange, due to the faster spatial decay of the second-order interactions ($1/r^6$ instead of $1/r^3$).

In the case of the two-atom Ramsey experiment [Fig.~\ref{fig:fig4}(b)], the oscillation frequency results from the combined effects of all van der Waals terms~\footnote{Here, we do not need to tune the microwave detuning to compensate the phase accumulation of the states $\ket{S,S}$ and $\ket{S',S'}$, since they have very similar van der Waals interactions.}: the calculated values $V_{SS'} = 2\pi \times 315$~kHz, $V_{SS} = 2\pi \times 102$~kHz and $V_{S'S'} = 2\pi \times 125$~kHz for $\theta=90$° lead to $\tilde{J}_{SS'} = J_{SS'} + V_{SS'} - (V_{SS} + V_{S'S'})/2 = 2\pi \times 500$~kHz, which is consistent with the experimental data.

We use the spin exchange method to measure the dependence of $|J_{SS'}|$ with~$r$ and~$\theta$. In Fig.~\ref{fig:fig4}(c), we recover the expected power law dependence $|J_{SS'}| \sim 1/r^6$. In Fig.~\ref{fig:fig4}(d), we measure the angular dependence of $|J_{SS'}|$ at~$r=9.9$~µm and find weak variations, that are compatible with the theory calculations at $B=46$~G. The weak angular dependence of $|J_{SS'}|$ originates from averaging over many virtual transitions to pair states of the form $\ket{P',P''}$ [Fig.~\ref{fig:fig1}(c)], with various angular dependences; contrary to the direct dipole-dipole interaction, the angular dependence of $|J_{SS'}|$ can be modified by the magnitude of the magnetic field, which affects the energy differences between the states and thus slightly changes the weight of the contribution of each state $\ket{P',P''}$~\cite{deLeseleuc_Weber_2018,Wadenpfuhl2024}.

\section{A spin exchange made resonant by local addressing}

\begin{figure}%[t]
\centering
\includegraphics{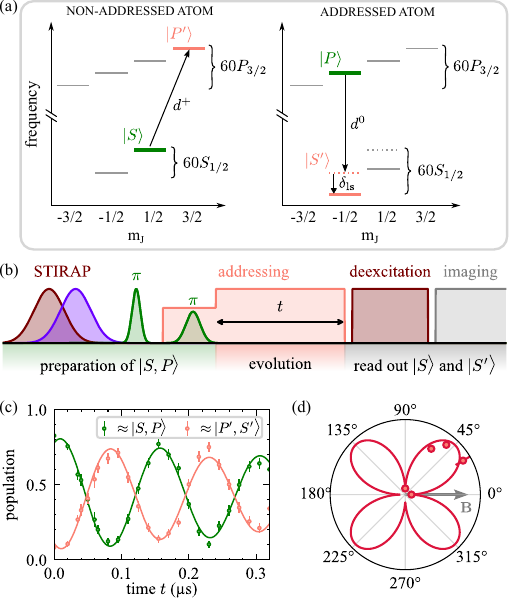}
\caption{{\bf Dipolar spin exchange involving four Rydberg levels.} (a)~Involved energy levels. (b)~Experimental sequence. (c)~Time evolution of the population of the two pair states $\ket{S,P}$ and $\ket{P',S'}$ when the interaction is tuned on resonance, for $r=9.9$~µm and $\theta=50$°. Solid lines are fits by damped sines. (d)~Measured (points) and calculated (lines) angular dependence of the interaction at $r=9.9$~µm.}
\label{fig:fig5}
\end{figure}

In the direct spin exchange interaction measured in Fig.~\ref{fig:fig2}, the pair states $\ket{S,P}$ and $\ket{P,S}$ are naturally degenerate. However, in the presence of a local light field acting on one of the two atoms, other pair states (involving Zeeman sublevels from the same $S,P$ manifold) can be brought into resonance. The resulting exchange is akin to Förster resonances~\cite{Ravets2014,Ravets2015}, but it is controlled with a light field rather than an electric field, which has the advantage of being easier to address spatially. In this section, we show proof-of-principle measurements of the coupling between $\ket{S,P}$ and $\ket{P',S'}$ involving the following four Rydberg levels: $\ket{S} = \ket{60S_{1/2},m_J=1/2}$, $\ket{S'} = \ket{60S_{1/2},m_J=-1/2}$, $\ket{P} = \ket{60P_{3/2},m_J=-1/2}$ and $\ket{P'} = \ket{60P_{3/2},m_J=3/2}$, as illustrated in Fig.~\ref{fig:fig5}(a). Due to the Zeeman shifts at $B=46$~G, the pair states $\ket{S,P}$ and $\ket{P',S'}$ are separated by~$42$~MHz, but they can be made resonant using a local light shift $\delta_\mathrm{ls}$ on the second atom that shifts down the energy of $\ket{S'}$ by $\delta_\mathrm{ls} \approx 2\pi\times 42$~MHz [Fig.~\ref{fig:fig5}(a)]. This local light shift is generated with the same addressing beam at 1014~nm that we use for local rotations~\cite{Bornet2024}.

The dipolar coupling $J_{SP,P'S'} = \bra{S,P} V_\mathrm{dd} \ket{P',S'}$ can be calculated by writing the dipolar interaction [Eq.~(\ref{eq:vdd})] using the spherical coordinates:
$\mathbf{d}_j = (d_j^- - d_j^+)/ \sqrt{2} \mathbf{e}_z + i (d_j^- + d_j^+) / \sqrt{2} \mathbf{e}_x + d_j^0 \mathbf{e}_y$, where the quantization axis is along $y$. We obtain
\begin{align}
	& V_\mathrm{dd} = \frac{1}{4\pi\epsilon_0 r^3} 
			\Bigg[ \frac{1-3\cos^2(\theta)}{2} \left(d_1^+ d_2^- + d_1^- d_2^+ + 2 d_1^0 d_2^0\right) \notag \\
			+ & \textstyle{\frac{3}{2\sqrt{2}}} \sin(2\theta) \left(e^{-i\phi} d_1^+ d_2^0 - e^{i\phi} d_1^- d_2^0 + e^{-i\phi} d_1^0 d_2^+ - e^{i\phi} d_1^0 d_2^- \right) \notag \\
			- & \textstyle{\frac{3}{2}} \sin^2(\theta) \left(e^{-2i\phi} d_1^+ d_2^+ + e^{2i\phi} d_1^- d_2^-\right) \Bigg]
			\label{eq:vdd_spherical}
\end{align}
where $\theta$ and $\phi$ are the spherical coordinates of the interatomic axis $\mathbf{e}_r = \sin(\theta)\cos(\phi) \mathbf{e}_z + \sin(\theta)\sin(\phi) \mathbf{e}_x + \cos(\theta) \mathbf{e}_y$. In Eq.~(\ref{eq:vdd_spherical}), the first line corresponds to the well-known~$1-3\cos^2(\theta)$ angular dependence measured in Fig.~\ref{fig:fig3}(b), whereas the $J_{SP,P'S'}$ coupling originates from the second line ($d_1^+ d_2^0$ term), with an expected $\sin(2\theta)$ angular dependence, together with a phase factor depending on~$\phi$. Contributions of this term have already been observed in Rydberg dressing experiments~\cite{Steinert2023}.

To observe a coherent exchange between $\ket{S,P}$ and $\ket{P',S'}$, we thus apply the sequence shown in Fig.~\ref{fig:fig5}(b). It is the same as the one of a usual spin exchange, except that the addressing lightshift is kept on during the whole time evolution. We use a different value of the addressing lightshift for the preparation and for the evolution, so that we limit the effect of interactions during the preparation. At the end of the sequence, the deexcitation pulse affects both $\ket{S}$ and $\ket{S'}$, so that the recapture probability is the sum of populations in $\ket{S}$ and $\ket{S'}$ (up to detection errors). This means that we do not distinguish an atom in $\ket{S}$ from an atom in $\ket{S'}$, but this limitation does not affect this proof-of-principle measurement.

As shown in Fig.~\ref{fig:fig5}(c) for $r = 9.9$~µm and $\theta=50$°, we obtain a coherent oscillation between the populations in $\ket{S,P}$ and $\ket{P',S'}$~\footnote{The mapping of the recapture data to the population in the target states $\ket{S,P}$ and $\ket{P',S'}$ is only approximate, due to the fact that we cannot distinguish all four states from each other: we only resolve whether the atom is in an $S$ state or a $P$ state. For example, the probability of the event \quota{atom 1 is recaptured and atom 2 is lost} is, up to detection errors, the sum of the populations in the states $\ket{S,P}$, $\ket{S',P}$, $\ket{S,P'}$ and $\ket{S',P'}$. But we assume that the three last states have a negligible population, since they are associated to off-resonant couplings.}. A fit by a damped sine wave (solid lines) gives $|J_{SP,P'S'}| = 2\pi \times (3.4 \pm 0.1)$~MHz. We repeat this measurement for various angles $\theta$ and obtain a reasonable agreement with the theory predictions [Fig.~\ref{fig:fig5}(d)]. This experiment is, to our knowledge, the first measurement of a coherent $d^+ d^0$ spin-exchange with Rydberg atoms, and may find applications for the quantum simulation of spin models with four (or more) spin states per atom.

\section{Outlook}
In this work, we have illustrated, in the simplest setting of just two atoms at a controlled distance, how various types of spin-exchange couplings can be implemented by encoding two spin states in Rydberg levels. Our careful benchmarking of the observed coherent interactions highlights the effects of positional disorder arising from the motion of the atoms, and that limits coherence to 10-20 interaction times. A possible way to mitigate these detrimental effects is to increase the interparticle distance~$r$ and use Rydberg states with a larger principal quantum number, such that the relative fluctuations of interaction energy are reduced as $\Delta J / |J| \propto \Delta r / r$. Another solution consists in trapping the Rydberg states using ponderomotive potentials~\cite{Barredo2020}, which can eliminate the effect of positional disorder when the trapping frequency~$\omega_\perp$ dominates over the fluctuations of interaction energy~$\Delta J$~\cite{Mehaignerie2023} (see also App.~\ref{App:positional_disorder}).
 
In the experiments reported here, the motion of the atoms leads to dephasing of the spin-exchange oscillations but, since we restrict ourselves to relatively short times, the dipolar interaction does not influence the motion of the atoms (as discussed in App.~\ref{App:vdW_forces}). However, going one step further, one can also take into account the mechanical effect of van der Waals or dipolar interactions on the motion of the atoms. This gives rise to a spin-motion coupling, observed for instance in~\cite{Chew_2022,Bharti2024}, that yields entanglement between internal and external degrees of freedom. From the perspective of quantum simulation of spin models, this is yet another decoherence channel; however, this spin-motion coupling may also open interesting perspectives, for example as a resource for generating entangled states~\cite{Mazza2020,Mehaignerie2023,Magoni2023,Wuster_2018}.
%For example, considering that the relative atomic motion is described by a one dimensional harmonic oscillator with annihilation and creation operators $\hat a, \hat a^\dagger$: $\hat r=r_0+\Delta x (\hat a+\hat a^\dagger)$, the Taylor expansion of Eq.~(\ref{eq:vdd}) contains a term of the form $(|J|\Delta x/r)(\hat a+\hat a^\dagger)(\hat \sigma_1^+\hat \sigma_2^- +\hat \sigma_1^-\hat \sigma_2^+)$ (with $\hat \sigma^\pm$ the pseudo-spin operators), which could be tuned.
It has also been suggested to implement GKP codes~\cite{Bohnmann2024} for quantum error correction.

\begin{acknowledgments}
We thank Johannes Mögerle for useful insights regarding the use of the Pairinteraction software package~\cite{Weber2017}. This work is supported by the Agence Nationale de la Recherche (ANR-22-PETQ-0004 France 2030, project QuBitAF), the European Research Council (Advanced Grant No. 101018511-ATARAXIA), and the Horizon Europe programme HORIZON-CL4-2022-QUANTUM- 02-SGA (Project No. 101113690 PASQuanS2.1).
\end{acknowledgments}

\appendix

\section{Condition to neglect the dipole-dipole forces} \label{App:vdW_forces}
Assuming that the atoms are in free flight amounts to neglecting the forces induced by the dipole-dipole interaction. This assumption is only valid at short times, when the dipole-dipole force has not yet affected the atomic positions. The time at which this assumption breaks down can be estimated from classical arguments. In the case of Fig.~\ref{fig:fig2}, the dominant potential is $\hbar J_{SP}(r)$, which gives rise to a force
\begin{align}
	F(r) &= - \hbar \frac{d J_{SP}}{dr} (r) \notag \\
	&= \frac{3 \hbar J_0 r_0^3}{r^4}
\end{align}
where $J_0 = J_{SP}(r_0)$. The classical equation of motion for the interatomic distance is
\begin{align}
	\mu \frac{d^2 r}{dt^2} = F(r)
\end{align}
with $\mu = m/2$ the reduced mass of the two-atom system. If we approximate the force by its value at the initial position~$r_0$ ($F(r) \simeq F(r_0)$), we obtain the displacement
\begin{align}
	r(t) - r_0 \simeq \frac{3 \hbar J_0}{2 \mu r_0} t^2.
\end{align}
The time at which this displacement gets on the same order of magnitude as the initial positional disorder ($\Delta r \simeq \sqrt{2}\sigma_\perp$) is given by
\begin{align}
	t \simeq \sqrt{\frac{2\sqrt{2}\mu r_0 \sigma_\perp}{3 \hbar J_0}}
\end{align}
With the parameters of Fig.~\ref{fig:fig2}(b), we find $t\simeq 14$~µs, which is indeed longer than our experimental durations.

\begin{figure*}%[t]
	\centering
	\includegraphics{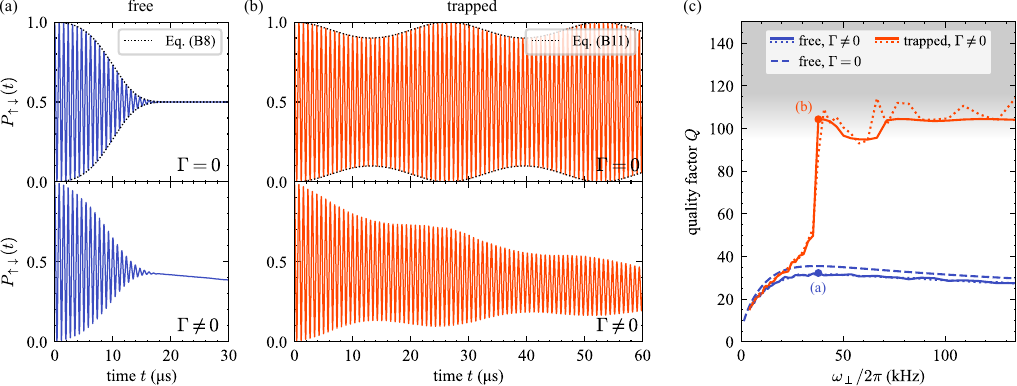}
	\caption{{\bf Roles of positional disorder and finite Rydberg lifetimes in the damping of a spin-exchange oscillation} for the experimental parameters of Fig.~\ref{fig:fig2}(c), that is to say $r_0=17$~µm, $\theta=90$°, $\ket{\uparrow} = \ket{60S_{1/2},m_J=1/2}$ and $\ket{\downarrow} = \ket{60P_{3/2},m_J=1/2}$, leading to an average interaction energy $J_0 = 2\pi \times 648$~kHz, $\alpha=3$ and negligible van der Waals interactions. We consider the ideal case of two atoms in the motional ground states of harmonic traps ($\bar{n}_\perp=0$), without state preparation errors.
		(a)~Simulated spin-exchange oscillations for atoms in free flight, where the trapping potential is switched off during the exchange, for an initial trapping frequency $\omega_\perp = 2\pi \times 38$~kHz. Top panel: infinite lifetime ($\Gamma =0$), with the analytical envelope of Eq.~(\ref{eq:spin_ex_free}) as a dotted line). Bottom panel: finite lifetime ($\Gamma \neq 0$). %The $y$-axis is what the experiment would measure for the probability~$P_{SP}(t)$ of the pair state~$\ket{S,P}$ based on the recapture probability.
		%Atomic motion is either modeled by classical trajectories (solid lines) or by a quantum treatment (dotted lines).
		(b)~Same in the case of trapped atoms. The analytical envelope is that of Eq.~(\ref{eq:spin_ex_trapped}).
		(c)~Quality factor as a function of the initial trapping frequency~$\omega_\perp$. Blue lines indicate the case of atoms in free flight, and orange lines display the case of trapped atoms. Solid lines are simulations with finite lifetimes assuming classical motion; dotted lines are the same simulations with quantum motion; and dashed lines are the case of infinite Rydberg lifetimes. %As a reference, the gray dashed line indicates the theory expectation for fixed atoms [Eq.~(\ref{eq:spin_ex_no_motion})].
		The gray region indicates the inaccessible range of quality factors due to finite Rydberg lifetimes. The points show the parameters used in panels~(a,b).
		}
	\label{fig:fig_appendix}
\end{figure*}

\section{A simple analytic model for the damping of a spin exchange oscillation} \label{App:positional_disorder}

We consider a system of two atoms whose internal degrees of freedom are restricted to the two pair states $\ket{\uparrow,\downarrow}$ and $\ket{\downarrow,\uparrow}$ with $\ket{\uparrow} = \ket{S}$ and $\ket{\downarrow} \in \{\ket{P}, \ket{S'}\}$. We model the atomic motion by classical trajectories that are independent from the internal state. This amounts to ignoring the possibility of entanglement between internal and external degrees of freedom; in particular, we neglect the state-dependent dipole-dipole forces which lead to additional displacements at late times (see App.~\ref{App:vdW_forces}). This approximation will be checked in App.~\ref{App:quantum_VS_classical_fluctuations}.
As positional disorder along~$z$ has a negligible effect on the dispersion of interaction energies for our parameters, we restrict the motion to one radial dimension with an initial trapping frequency $\omega_\perp$. We first sample an initial position $x_0$ and velocity $v_0$ from Gaussian distributions $\rho_r(x_0) = \exp(-x_0^2/2\Delta r^2) / \sqrt{2\pi} \Delta r$ and $\rho_v(v_0) = \exp(-v_0^2/2\Delta v^2) / \sqrt{2\pi} \Delta v$. Although the motion is classical, we use the standard deviations of positions and velocities of a quantum harmonic oscillator: $\Delta r = \sqrt{2} \sigma_\perp = \sqrt{\hbar ( 2\bar{n}_\perp + 1) / m \omega_\perp}$ and $\Delta v = \omega_\perp \Delta r$. Then we make the interatomic distance~$r$ evolve according to
\begin{align}
	\label{eq:classical_trajectories}
	r(t,x_0,&v_0) = \\
	r_0 \; + &\left\{
	\begin{array}{ll}
		x_0 + v_0 t \quad \text{for atoms in free flight} \\ [6pt]
		x_0 \cos(\omega_\perp t) + \frac{v_0}{\omega_\perp}  \sin(\omega_\perp t) \quad \text{for trapped atoms}.
	\end{array} \notag
	\right.
\end{align}
with $r_0$ the average distance.

Each realization of the positional disorder $(x_0,v_0)$ corresponds to a given dynamics for the internal state $\ket{\psi(t,x_0,v_0)}$, which is governed by the Schrödinger equation:
\begin{align}
	i\hbar \frac{d\ket{\psi}}{dt}  = H\left[r(t,x_0,v_0)\right] \ket{\psi} \ ,
	\label{eq:schrodinger_eq}
\end{align}
where the Hamiltonian for the internal state is given by
\begin{align}
	H(r) = \hbar
	\begin{bmatrix}
		0 & J(r) \\
		J(r) & 0
	\end{bmatrix}
	\label{eq:hamiltonian_internal}
\end{align}
in the basis $\{ \ket{\uparrow,\downarrow}, \ket{\downarrow,\uparrow}\}$.
The spin exchange coupling can be written $J(r) = J_0 \left(r_0 / r\right)^\alpha$ with $\alpha = 3$ (resp. $\alpha = 6$) for a direct (resp. indirect) spin-exchange. Note that compared with Hamiltonian~(\ref{eq:H_eff_two_levels}) of the main text, we have neglected the diagonal van der Waals term, which will only lead to a global phase accumulation.
Writing $\ket{\psi} = c_+ \ket{+} + c_- \ket{-}$ with $\ket{\pm} = (\ket{\uparrow,\downarrow} \pm \ket{\downarrow,\uparrow})/\sqrt{2}$, Eq.~(\ref{eq:schrodinger_eq}) becomes
\begin{align}
	\frac{d c_\pm}{d t}  = \mp i J\left[r(t,x_0,v_0)\right] c_\pm
\end{align}
whose solution is
\begin{align}
	c_\pm(t,x_0,v_0) = \frac{1}{\sqrt{2}} \exp \left(\mp i \int_0^t J\left[r(\tau,x_0,v_0)\right] d\tau \right)
\end{align}
for the initial state of a spin exchange~$\ket{\psi(t=0)} = \ket{\uparrow,\downarrow}$.
Each realization of the positional disorder leads to the following probability for the state~$\ket{\uparrow, \downarrow}$:
\begin{align}
	\label{eq:spin_ex_for_r0v0}
	p_{\uparrow,\downarrow} (t,x_0,v_0) &= \left|\langle \uparrow,\downarrow \ket{\psi(t,x_0,v_0)}\right|^2 \\
	&= \frac{1}{2} \left|c_+ (t,x_0,v_0) + c_- (t,x_0,v_0) \right|^2 \notag \\
	&= \frac{1}{2} \left[1 + \cos\left(2 \int_0^t J\left[r(\tau,x_0,v_0)\right] d\tau\right)\right] \notag
\end{align}
%One recovers the ideal spin-exchange oscillation $p_{\uparrow,\downarrow} = \frac{1}{2} \left[1 + \cos\left(2 J_0 t\right)\right]$ if $J(r) = J_0$.
Finally, our experiment measures the probability of the state~$\ket{\uparrow, \downarrow}$ averaged over all realizations of the positional disorder:
\begin{align}
	P_{\uparrow,\downarrow} (t) &= \int dx_0 \int dv_0 \; \rho_r(x_0) \rho_v(v_0) \; p_{\uparrow,\downarrow} (t,x_0,v_0) \ .
	\label{eq:spin_ex_general}
\end{align}

We now evaluate Eq.~(\ref{eq:spin_ex_general}) in the case of (i)~atoms in free flight and (ii)~trapped atoms, following the classical trajectories given by Eq.~(\ref{eq:classical_trajectories}). To do so, we linearize~$J(r)$ around~$r_0$: $J(r) \approx J_0 - \alpha J_0 (r-r_0)/r_0$ which is valid if $|r-r_0| \ll r_0$. 
The Gaussian distribution of distances translates into a Gaussian distribution of interaction energies with standard deviation~$\Delta J = \alpha J_0 \Delta r / r_0$, allowing us to use Gaussian integration formulas.

(i)~For atoms in free flight, we obtain
\begin{align}
	P_{\uparrow,\downarrow}^\text{free} (t) = \frac{1}{2} \left[1 +  e^{-2 \Delta J^2 t^2 \, \left(1 + \frac{1}{4} \omega_\perp^2 t^2\right)} \cos\left(2 J_0 t\right) \right].
	\label{eq:spin_ex_free}
\end{align}
The amplitude of the oscillation is reduced by an exponential envelope which contains a quadratic term~$\propto t^2$ due to the initial position dispersion, and a quartic term~$\propto t^4$ which originates from the initial velocity dispersion.
In Fig.~\ref{fig:fig_appendix}(a, top panel), we compare the theory profile of Eq.~(\ref{eq:spin_ex_free}) to a numerical simulation for the experimental parameters of Fig.~\ref{fig:fig2}(c), assuming that atoms are perfectly cooled to the motional ground state. We obtain a perfect agreement, thus validating the approximations which were made in the derivation, namely: neglecting positional disorder along~$z$ and linearizing the dipole-dipole potential. We also include finite Rydberg lifetimes in the simulation (bottom panel) and conclude that positional disorder is the main limitation for atoms in free flight in this range of parameters.

%At short times $t \ll 1/\omega_\perp$, the spin exchange oscillation~(\ref{eq:spin_ex_free}) can be simplified as
%\begin{align}
%	P_{\uparrow,\downarrow}^\text{fixed} (t) = \frac{1}{2} \left[1 +  e^{-2 \Delta J^2 t^2} \cos\left(2 J_0 t\right) \right]
%	\label{eq:spin_ex_no_motion}
%\end{align}
%which describes the case of fixed atoms (no velocities).

The $1/\sqrt{e}$ damping time in Eq.~(\ref{eq:spin_ex_free}) is given by
\begin{align}
	\tau &= \frac{\sqrt{2}}{\omega_\perp} \sqrt{\sqrt{1+\frac{\omega_\perp^2}{4 \Delta J^2}}-1}
	\label{eq:damping_time}
\end{align}
The dependence of the quality factor $Q = J_0 \tau$ on the initial trapping frequency $\omega_\perp$ is plotted as a dashed blue line in Fig.~\ref{fig:fig_appendix}(c) for the parameters of Fig.~\ref{fig:fig2}(c) and for $\bar{n}_\perp=0$.
Using the fact that $\Delta J$ depends on $\omega_\perp$ as $\Delta J = \alpha J_0 \Delta r(\omega_\perp) / r_0$, we find that the quality factor is maximized for an optimal trapping frequency
\begin{align}
	\omega_\perp^\mathrm{opt} = 2 \left[\frac{4 \alpha^2 \hbar (2\bar{n}_\perp+1) J_0^2}{m r_0^2}\right]^{1/3},
\end{align}
corresponding to $\tau(\omega_\perp^\mathrm{opt}) = 2 / \omega_\perp^\mathrm{opt}$. We find $\omega_\perp^\mathrm{opt} = 2\pi \times 36$~kHz and $Q(\omega_\perp^\mathrm{opt}) = 34$. The dominant contribution to the damping is the initial position dispersion for $\omega_\perp < \omega_\perp^\mathrm{opt}$ and the initial velocity dispersion for $\omega_\perp > \omega_\perp^\mathrm{opt}$. Numerical simulations in the presence of finite Rydberg lifetimes [solid blue lines in Fig.~\ref{fig:fig_appendix}(c)] show that the quality factor is weakly affected by the lifetimes in this range of parameters.

(ii)~For trapped atoms, we obtain
\begin{align}
	P_{\uparrow,\downarrow}^\text{trapped} (t) = \frac{1}{2} \left[ 1 + e^{-4 \left(\frac{\Delta J}{\omega_\perp}\right)^2 \left(1-\cos(\omega_\perp t)\right)} \cos\left(2 J_0 t\right) \right].
	\label{eq:spin_ex_trapped}
\end{align}
The envelope shows periodic revivals due to the oscillations of the atomic positions in the traps, which cause the probability of each classical trajectory [Eq.~(\ref{eq:spin_ex_for_r0v0})] to rephase. Those revivals were also derived in Ref.~\cite{Mehaignerie2023} using a full quantum treatment of the atomic motion (see Appendix~\ref{App:quantum_VS_classical_fluctuations}).
%A perturbative expansion of $\cos(\omega_\perp t)$ at short times shows that the early dynamics of trapped atoms [Eq.~(\ref{eq:spin_ex_trapped})] is also equivalent to fixed atoms [Eq.~(\ref{eq:spin_ex_no_motion})].
The modulation of the envelope in Eq.~(\ref{eq:spin_ex_trapped}) can be suppressed by increasing the trapping frequency~$\omega_\perp$ so to reach the regime $\omega_\perp \gg \Delta J$.
Using the dependence of $\Delta J$ on $\omega_\perp$, this condition becomes $\omega_\perp \gg \omega_\perp^\mathrm{opt} / 3$.

In Fig.~\ref{fig:fig_appendix}(b, top panel), we simulate the spin exchange of two trapped atoms in their motional ground state for $\omega_\perp \simeq \omega_\perp^\mathrm{opt}$ and find again a good agreement with the predicted profile of Eq.~(\ref{eq:spin_ex_trapped}). Now finite Rydberg lifetimes become the dominating limiting factor to the damping [Fig.~\ref{fig:fig_appendix}(b, bottom panel)], leading to a significant improvement in the coherence of the oscillation compared with the case of free atoms.

To estimate by how much Rydberg trapping can improve the damping, we repeat the simulations for various initial trapping frequencies~$\omega_\perp$ and extract the quality factor of each resulting spin-exchange oscillation. Due to the revivals, fitting $P_{\uparrow \downarrow}(t)$ by a Gaussian envelope is not reliable, so we define the damping time~$\tau$ as the $1/\sqrt{e}$ decay time for the contrast of $P_{\uparrow \downarrow}(t)$. The results are shown in Fig.~\ref{fig:fig_appendix}(c) as orange lines: the quality factor reaches $\sim100$ for trapped Rydberg atoms and saturates starting from $\omega_\perp \sim \omega_\perp^\mathrm{opt}$ due to the finite lifetimes.
%As a reference, we also show the theoretical case of fixed atoms with infinite lifetimes [Eq.~(\ref{eq:spin_ex_no_motion})]. It is larger than the quality factor for atoms in free flight due to the harmful effect of the initial velocity dispersion at large~$\omega_\perp$; but it is smaller than the one of trapped Rydberg atoms due to the beneficial effect of revivals.

\section{Quantum treatment of positional disorder} \label{App:quantum_VS_classical_fluctuations}
Numerical simulations performed throughout the paper are Monte-Carlo simulations where atomic positions and velocities are sampled from Gaussian distributions and evolved according to classical trajectories (similarly to Refs.~\cite{Chew_2022,bao_dipolar_2023,Holland_2023,Kim_Heisenberg_2024}). One may wonder whether the results remain valid as the atomic wavefunctions are approaching the ground state of the trap. This case is treated in details in Ref.~\cite{Mehaignerie2023,Picard_2025}, here we simply implement numerical simulations of the quantum equations of motion with our experimental parameters.

Restricting the internal degrees of freedom to $\ket{\uparrow,\downarrow}$ and $\ket{\downarrow,\uparrow}$, the two-atom system is described by the following relative wavefunction:
\begin{align}
	\langle r \ket{\psi_\mathrm{rel}(t)} = \varphi_{\uparrow,\downarrow}(r,t) \ket{\uparrow,\downarrow} + \varphi_{\downarrow,\uparrow}(r,t) \ket{\downarrow,\uparrow}
	\label{eq:def_wavefunction}
\end{align}
with the normalization condition $\int dr |\varphi_{\uparrow,\downarrow}(r,t)|^2 + \int dr|\varphi_{\downarrow,\uparrow}(r,t)|^2 = 1$. Here also, we have restricted the motion to one dimension, as the effect of transverse positional disorder is negligible for our parameters. The dynamics of the system is governed by the Schrödinger equation:
\begin{align}
	i\hbar \frac{d \ket{\psi_\mathrm{rel}}}{dt}  = \left(\frac{\hat{p}^2}{2\mu}
	+ \frac{1}{2} \mu \omega_\perp^2 (\hat{r}-r_0)^2
	+ H(\hat{r})\right) \ket{\psi_\mathrm{rel}}.
	\label{eq:schrodinger_eq_quantum}
\end{align}
where $\hat{r}$ is the distance operator, $r_0$ is the average distance, $\mu = m/2$ is the reduced mass of the two-atom system, and the effective Hamiltonian for the dipole-dipole interactions is given by
\begin{align}
	H(\hat{r}) = \hbar
	\begin{bmatrix}
		V(\hat{r}) & J(\hat{r}) \\
		J(\hat{r}) & V(\hat{r})
	\end{bmatrix}
	\label{eq:hamiltonian_dipole-dipole_for_app}
\end{align}
in the basis $\{ \ket{\uparrow,\downarrow}, \ket{\downarrow,\uparrow}\}$. Compared to Hamiltonian~(\ref{eq:hamiltonian_internal}), we include the diagonal term~$V$ as it can affect the motion of the wavepackets. Equation~(\ref{eq:schrodinger_eq_quantum}) can be rewritten for~$\varphi_\pm = \left(\varphi_{\uparrow,\downarrow} \pm \varphi_{\downarrow,\uparrow} \right)/\sqrt{2}$ as two independent differential equations:
\begin{align}
	\displaystyle \frac{\partial \varphi_\pm}{\partial t} = \displaystyle \frac{i \hbar}{2 \mu} \frac{\partial^2 \varphi_\pm}{\partial r^2} -i V_\pm(r) \varphi_\pm
	\label{eq:quantum_equations_of_motion_two_atoms}
\end{align}
where we have defined the state-dependent potential $V_\pm(r) = V(r) \pm J(r) + \frac{1}{2} \mu \omega_\perp^2 (r - r_0)^2$. An important consequence of this potential is the possibility of spin-motion coupling, meaning that atomic motion is affected by the internal state --- an effect which is not taken into account in the classical Monte-Carlo simulation.

We perform numerical simulations of the quantum equations of motion~(\ref{eq:quantum_equations_of_motion_two_atoms}) using a Crank–Nicolson scheme, starting from the state $\langle r \ket{\psi_\mathrm{rel}(t=0)} = \varphi_0(r) \ket{\downarrow,\uparrow}$
with $|\varphi_0(r)|^2 = \exp\left[-(r-r_0)^2/2\Delta r^2 \right] / \sqrt{2\pi}\Delta r$ the distribution of interatomic distances in the ground state of the trap at frequency~$\omega_\perp$. We then evaluate the probability $P_{\uparrow\downarrow}(t) = \int dr |\langle r, \uparrow, \downarrow \ket{\psi_\mathrm{rel}(t)}|^2$. We also add the finite Rydberg lifetimes as an additional decoherence channel using a quantum Monte-Carlo method~\cite{Dalibard_1992}.

The results are shown as dotted lines in Fig.~\ref{fig:fig_appendix}(c). We find that they give the same damping as the classical Monte-Carlo simulations ---we also checked that the probabilities $P_{\uparrow\downarrow}(t)$ from both simulation methods overlap perfectly in the time domain---, confirming that for our range of parameters, quantum fluctuations of positions and velocities can be safely replaced by classical fluctuations, and that spin-motion coupling does not affect the damping of the oscillation.
%However it was shown in Refs.~\cite{Sommer_2016,Bharti2024} that in the context of optical lattices, signatures of spin-motion coupling can be found in the coherence of Ramsey oscillations.

\bibliography{refs}

\end{document}